\documentclass[conference]{IEEEtran}
\ifCLASSINFOpdf
\else
\fi
\usepackage{graphicx}
\usepackage{lettrine}
\usepackage{setspace}
\usepackage{amsmath}
\usepackage{amssymb} 
\usepackage{caption}
\usepackage{subcaption}
\usepackage{bm} 
\usepackage{cite}
\usepackage{float}
\usepackage[usenames,dvipsnames]{pstricks}
\usepackage{epsfig}
\usepackage{pst-grad} 
\usepackage{pst-plot} 
\usepackage{tabularx}
\usepackage{flushend}

\usepackage[T1]{fontenc}
\usepackage{soul}
\newcommand{\lb} {\left}
\newcommand{\rb} {\right}
\newcommand{\nn} {\nonumber}
\usepackage{xcolor}
\usepackage{lipsum}

\allowdisplaybreaks

\allowdisplaybreaks
\flushend
\usepackage[utf8]{inputenc}

 \IEEEaftertitletext{\vspace{-3\baselineskip}}

\begin{document}

\title{Secrecy Performance of a Keyhole-based Multi-user  System with Multiple Eavesdroppers}

\author{
    \IEEEauthorblockN{Parwez Alam\IEEEauthorrefmark{1}, 
    Ankit Dubey\IEEEauthorrefmark{2}, 
    Jules M. Moualeu\IEEEauthorrefmark{3}, 
    Telex M. N. Ngatched\IEEEauthorrefmark{4},
    and Chinmoy Kundu\IEEEauthorrefmark{5},
    }
    \IEEEauthorblockA{\IEEEauthorrefmark{1}\IEEEauthorrefmark{2}Department of EE, Indian Institute of Technology Jammu, Jammu \& Kashmir, India}
    \IEEEauthorblockA{\IEEEauthorrefmark{3}School of Electrical and Information Engineering, University of the Witwatersrand, Johannesburg, South Africa}
    \IEEEauthorblockA{\IEEEauthorrefmark{4}Department of Electrical and Computer Engineering, McMaster University, Hamilton ON, Canada}
    \IEEEauthorblockA{\IEEEauthorrefmark{5}Wireless Communications Laboratory, Tyndall National Institute, University College Cork, Cork, Ireland}
    
    \textrm{\{\IEEEauthorrefmark{1}2022ree0013},{\IEEEauthorrefmark{2}ankit.dubey}\}@iitjammu.ac.in, \IEEEauthorrefmark{3}jules.moualeu@wits.ac.za,
    \IEEEauthorrefmark{4}ngatchet@mcmaster.ca,
    \IEEEauthorrefmark{5}chinmoy.kundu@tyndall.ie
     }

\maketitle
 \thispagestyle{empty}
\pagestyle{plain} 

\maketitle
\begin{abstract} 
This paper investigates the secrecy performance of a keyhole-aided multi-user communication network in the presence of multiple eavesdroppers. The communication happens through the same keyhole for legitimate users and eavesdroppers. In this context, the secrecy performance is evaluated for a user scheduling technique by obtaining the exact closed-form expression of secrecy outage probability (SOP). Further, a simplified asymptotic SOP expression is derived assuming high signal-to-noise ratio (SNR) scenario for a better understanding of the impact of system parameters. The effect of the keyhole parameters, number of users, number of eavesdroppers, and threshold secrecy rate on the SOP performance are also investigated for the considered system model. In the high-SNR regime, the asymptotic SOP saturates to a constant value and does not depend on the keyhole parameter and the channel parameter of the source-to-keyhole channel.
\end{abstract}
\vspace{-.2cm}
\section{Introduction}
In a highly scattered environment, single-user communication with multiple transmit and receive antennas can achieve high spectral efficiency \cite{Chizhik-2002_Keyholes_correlations}. Moreover, such an environment makes the signal from each transmit antenna appear highly uncorrelated at each receive antenna. If the signals arriving at the receive antennas are correlated, then the spectral efficiency is reduced. However, a low signal correlation at each of the receive antennas is not a guarantee to achieve high capacity. A degenerative propagation scenario called ``\textit{keyhole}'' may reduce capacity even though the system has uncorrelated transmit and receive signals \cite{chizhik-2000_capacities,gesbert_outdoor_2002, Loyka_2002_MIMO_Capacity}.

Realistically, a keyhole effect may arise during the propagation of electromagnetic radiation through a hallway, tunnel, or corridor where the signal from the transmitter can propagate to the receiver only through the keyhole. A portion of the incident electric field is transmitted through the keyhole. The ratio of the transmitted field through the keyhole to the incident field on the keyhole is called the scattering cross-section of the keyhole \cite{Chizhik-2002_Keyholes_correlations}. The dimensions of the keyhole should be much smaller than the corresponding wavelength \cite{Levin_2011_Multi_Keyholes}. Although the signals at the antennas are not correlated, the keyhole effect reduces the multiple-input multiple-output (MIMO) channel capacity comparable to the single-input single-output (SISO) channel capacity \cite{Almers_Keyhole_Peter_2006, Loyka_2005_Multiantenna}. As the keyhole channel reduces the channel rank to one regardless of the number of antennas at the transmitter and receiver, the spatial diversity is more beneficial than the multiplexing gain of the MIMO channel \cite{Levin_2011_Multi_Keyholes}. 
Motivated by this, the authors in \cite{Sanayei_Shahab_2007} studied a transmit and receive antenna selection in the presence of keyhole condition wherein it was shown that the antenna selection did not decrease the diversity of the keyhole channel.

Due to the broadcast nature of wireless media, a wireless transmission is susceptible to eavesdropping.  Physical layer security (PLS) is emerging as an efficient solution to provide security in wireless networks due to its low-complexity techniques which use physical characteristics of the wireless channel to provide security \cite{Leung_1978_Gaussian_Wiretap_Channel}.
In \cite{Zang_2022_Secrecy} and \cite{Zang_2023_PHY_Security}, a keyhole-aided MIMO-vehicle-to-vehicle (MIMO-V2V) in the presence of an eavesdropper is investigated, assuming that both the main and eavesdropper channels are modeled as two mutually independent MIMO keyhole channels.
Specifically, the closed-form expressions of the non-zero secrecy capacity, secrecy outage probability (SOP), and average secrecy capacity (ASC)  are derived. In \cite{Sultana_2016_Security}, a secure communication scenario through multi-keyhole MIMO channels is considered where the effect of the number of keyholes on the ergodic secrecy capacity is investigated.

A keyhole channel resembles a cascade of two fading channels where local scatterers in the vicinity produce more than one multiplicative small-scale
fading processes, i.e., cascaded fading \cite{Levin_2011_Multi_Keyholes,Zhang_2022_MIMO-HARQ}. In \cite{tashman_cascaded_2021}, the PLS of a system consisting of a transmitter and receiver in a cascaded $\kappa$-$\mu$ channel is investigated in the presence of multiple colluding and non-colluding eavesdroppers. In \cite{Tashman_PLS_2020}, the securecy performance of a cognitive radio networks (CRNs) in a cascaded Rayleigh fading channel is investigated. A user scheduling approach is considered to secure the keyhole-based power line communication (PLC) in the presence of an eavesdropper in \cite{Kundu_2023_Destination-Scheduling}. 


MIMO transceivers employ multiple radio frequency (RF) chains and low-noise amplifiers (LNA), which is expensive. Furthermore, MIMO systems are also computationally expensive due to high-dimensional signaling and complex decoding algorithms \cite{Sanayei_Shahab_2007}. in the presence of a keyhole condition, spatial diversity is more beneficial than the multiplexing gain, which negates the application of MIMO processing. On the other hand, a single-input single-output (SISO) system, which leverages a single antenna for transmission and reception of the RF signal, offers a cost-effective and power-efficient solution, particularly in scenarios where hardware complexity and power consumption pose significant challenges, such as in Internet of Things (IoT) networks deployed in smart cities, smart home, and smart industries \cite{Fuqaha_IOT_Survey_2015}. Thus, the security of future IoT-based multi-user systems in keyhole channels would require inexpensive technology.  However, the PLS techniques for multiple distributed low-complexity users in keyhole channels are not adequately studied. Moreover, most studies on the security of keyhole channel employ MIMO transceivers which is expensive \cite{Zang_2022_Secrecy, Zang_2023_PHY_Security, Sultana_2016_Security}. Assuming cascaded fading channels, no work in the literature has considered the detrimental effect of multiple eavesdroppers (see  \cite{tashman_cascaded_2021, Tashman_PLS_2020, Kundu_2023_Destination-Scheduling} and the references therein). Moreover, the works in  \cite{tashman_cascaded_2021} and \cite{Tashman_PLS_2020} do not consider user scheduling techniques for security enhancement. Though the work in \cite{Kundu_2023_Destination-Scheduling} adopts a user scheduling technique in the presence of a single eavesdropper, a PLC channel is assumed, which is quite different from the wireless channel.
 
Motivated by the above discussion, the present work considers a system where a source is communicating with multiple users via a keyhole channel in the presence of multiple eavesdroppers. The communication occurs through the same keyhole for all the users and eavesdroppers. In addition, a user scheduling technique is proposed for security enhancement, which inherently has less complexity and can be deployed in future low-cost low-complexity networks.  Our key contributions are as follows.

 
\begin{itemize}


\item We derive the closed-form expression for the SOP of the keyhole channel in the presence of multiple eavesdroppers where a user is selected for scheduling among multiple users. 

\item We derive a simplified asymptotic SOP for the high SNR approximation to gain deeper insights.

\item We evaluate the impact of scattering cross-section of keyhole, number of users and eavesdroppers,  and the secrecy rate threshold on the system secrecy performance.
     
\end{itemize}

The rest of the paper is organized as follows. Section \ref{sec_system_model} introduces the system model of the proposed keyhole-based multi-user multi-eavesdropper system. Section \ref{sec_SOP} presents the theoretical framework of the SOP for the underlying system, while Section \ref{sec_ASYM_SOP} highlights the high-SNR asymptotic analysis of the SOP. In Section \ref{sec_results}, numerical results are presented. Finally, concluding remarks are provided in Section \ref{sec_conclusion}.

\textit{Notation:} $\mathbb{P}[\cdot]$ represents the probability of the occurrence of an event, $\mathbb{E}[\cdot]$ is the expectation operator, $K_1\lb(\cdot\rb)$ denotes the modified Bessel function of the second kind of order 1. The probability density function (PDF) and the cumulative distribution function (CDF) of a random variable $X$ are denoted by $f_{X}(\cdot)$ and $F_{X}(\cdot)$, respectively.


\section{SYSTEM MODEL}
\label{sec_system_model}
\begin{figure}
 \centering
\includegraphics[width=2.8in]{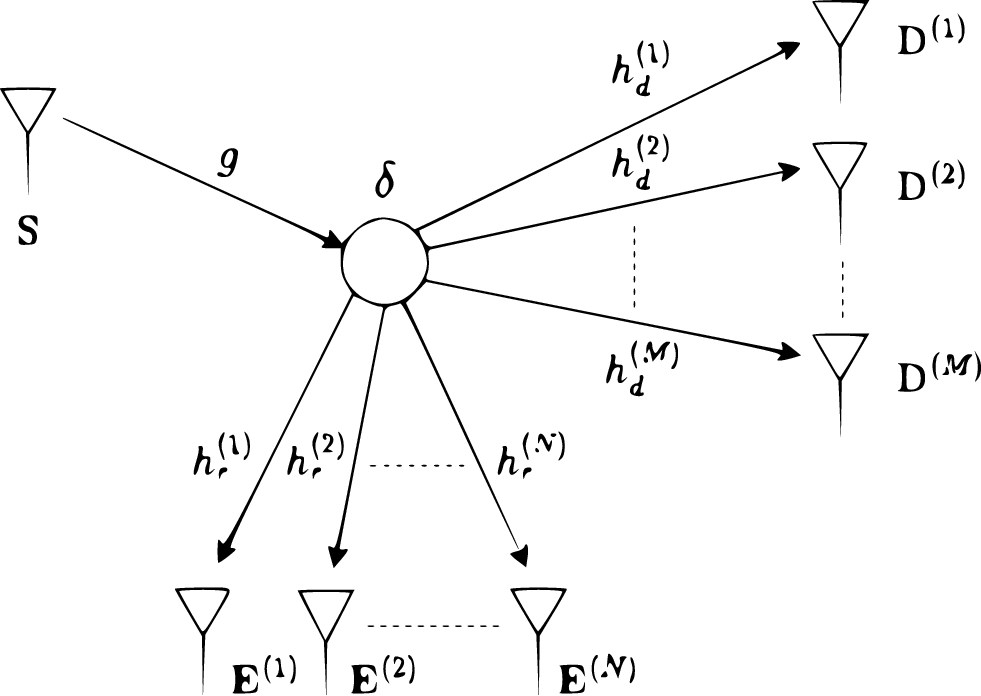} 
 \vspace{-.1cm}
 \caption{A keyhole-based system model with a source, multiple users, and multiple eavesdroppers.}
 \vspace{-.5cm}
 \label{fig_SM_PH2}
 \end{figure}
 
We consider a multi-user 
 communication system with multiple eavesdroppers where a source $\textrm{S}$ is communicating with $M$ legitimate users ${\textrm{D}}^{(m)}$ for $m\in \mathcal{M} =\{1,2,\dots, M\}$ in the presence of $N$ passive eavesdroppers  ${\textrm{E}}^{(n)}$ for $n\in \mathcal{N} =\{1,2,\dots, N\}$ as shown in Fig. \ref{fig_SM_PH2}. The communication takes place through a keyhole with scattering cross-section $\delta$, separating the regions containing the users and the source. The eavesdroppers reside in the same region containing the users. It is assumed that the direct links from S to $\textrm{D}^{(m)}$  or $\textrm{E}^{(n)}$ for any $m\in \mathcal{M}$ and $n\in \mathcal{N}$, respectively, does not exist. 
The  fading channels between the source-to-keyhole, keyhole-to-$m$-th destination, and keyhole-to-$n$-th eavesdropper are denoted by $g$, $h_d^{(m)}$, and $h_e^{(n)}$, respectively. 
We consider $g$, $h_d^{(m)}$ for all $m\in \mathcal{M}$, and $h_e^{(n)}$ for all $n\in \mathcal{N}$ to be mutually independent, complex Gaussian random variables (RV).  
We also assume that $h_d^{(m)}$ for all $m\in \mathcal{M}$ and $h_e^{(n)}$ for all $n\in \mathcal{N}$ are identically distributed. However, $g$, $h_d^{(m)}$ any $m$, and $h_e^{(n)}$ for any $n$ are not identically distributed. The magnitude of the links follows the Rayleigh distribution. Hence, the channel powers $|g|^2$, $|h_d^{(m)}|^2$ for any $m\in \mathcal{M}$, and $|h_e^{(n)}|^2$ for any $n\in \mathcal{N}$, follow an exponential distribution with parameters ${\zeta_g}$, ${\zeta_{h_d}}$ and ${\zeta_{h_e}}$, respectively.
The CDF and PDF of an exponentially distributed RV $Z$ with parameter $\zeta_{Z}$ can be expressed respectively as
\begin{align}
\label{cdf_Z}
    F_{Z} (z) &=1-\exp{\lb(-\frac{z}{\zeta_{Z}}\rb)}, \\
\label{pdf_Z}
  f_{Z} (z)&= \frac{1}{\zeta_{Z}}\exp{\lb(-\frac{z}{\zeta_{Z}}\rb)}.
\end{align}


The signal received at $\textrm{D}^{(m)}$ through the keyhole is given by
 \begin{align}
 \label{User_Rx_signal}
     y_d^{(m)} = g \delta h_d^{(m)} x + n_d^{(m)},
 \end{align}
where $x$ is the transmitted signal with average transmitted power $\mathbb{E}\lb({|x|^2}\rb)=P$ and  $n_d^{(m)}\sim \mathcal {CN}\left(0, \sigma_d^2\right)$ for each $m$, is the additive white Gaussian noise (AWGN) at  $\textrm{D}^{(m)}$ with mean zero and variance $\sigma_d^2$.
Similarly, the received signal at $\textrm{E}^{(n)}$ can be written as
\begin{align} 
\label{Evesd_Rx_signal}
    y_{e}^{(n)} = g \delta h_e^{(n)} x + n_{e}^{(n)},
\end{align}
where  $n_{e}^{(n)}\sim \mathcal {CN}\left (0, \sigma_e^2\right)$ for each $n$, is the AWGN at $\textrm{E}^{(n)}$ with mean zero and variance $\sigma_e^2$. 

To maximize the signal-to-noise ratio (SNR) at the users, we select the best user $m^*$ among $M$ users for scheduling, which has the maximum source-to-user channel power gain. In this case, the instantaneous SNR at the best user $\textrm{D}^{(m^*)}$ can be expressed as
\begin{align}
\label{inst_snr_D}
   \gamma_d = \bar{\gamma}_d|g|^2\delta^2 \max_{m\in \mathcal {M}}\{ |h_d^{(m)}|^2 \},
\end{align}
where $\bar{\gamma}_d = \frac{P}{\sigma_d^2}$. To measure the worst case secrecy, we measure the secrecy against the eavesdropper $n^*$ which has the best source-to-eavesdropper channel power gain. In this case, the instantaneous SNR at 
$\textrm{E}^{(n^*)}$ can be expressed as 
\begin{align}
\label{inst_snr_E}
   \gamma_e = \bar{\gamma}_e|g|^2 \delta^2 \max_{n\in \mathcal {N}}\{(|h_e^{(n)}|^2)\}, 
\end{align} 
where $\bar{\gamma}_e = \frac{P}{\sigma_e^2}$.


Using the order statistics, the CDF of $\bar{\gamma}_d \max_{\forall m}\{\lvert h_d^{(m)}\rvert^2\}$ can be expressed as
\begin{align}
F_X(x)&=\mathbb{P}\lb[\max_{\forall m}\{\lvert h_d^{(m)}\rvert^2\} \le \frac{x}{\bar{\gamma}_d}\rb] = \lb[1-\exp{\lb(-\frac{x}{\zeta_{h_d} \bar{\gamma}_d}\rb)}\rb]^M \nn\\
\label{cdf_X}
&=1+\sum^M_{m=1}(-1)^m\binom{M}{m}\exp{\lb({-\frac{mx}{\zeta_{h_d}\bar{\gamma}_d}}\rb)},
\end{align}
where $X=\bar{\gamma}_d \max_{\forall m}\{\lvert h_d^{(m)}\rvert^2\}$. By differentiating \eqref{cdf_X}, the corresponding PDF is obtained as
  \begin{align}
  \label{pdf_X}
f_{X}(x)
&=\sum^M_{m=1}(-1)^{m+1}\binom{M}{m}\frac{m}{\zeta_{h_d} \bar{\gamma}_d}\exp{\lb({-\frac{mx}{\zeta_{h_d}\bar{\gamma}_d}}\rb)}.
  \end{align}
Similarly, the CDF of $\bar{\gamma}_e \max_{\forall n}\{\lvert h_e^{(n)}\rvert^2\}$ can be expressed as
\begin{align}
\label{cdf_Y}
   F_Y{(y)}&=1+\sum^N_{n=1}(-1)^n\binom{N}{n}\exp{\lb({-\frac{ny}{\zeta_{h_e}\bar{\gamma}_e}}\rb)},
\end{align}
where $Y= \bar{\gamma}_e \max_{\forall n}\{\lvert h_e^{(n)}\rvert^2\}$. The corresponding PDF is obtained as 
\begin{align}
\label{pdf_Y}
    f_Y(y)=\sum^N_{n=1}(-1)^{n+1}\binom{N}{n}\frac{n}{\zeta_{h_e} \bar{\gamma}_e}\exp{\lb({-\frac{ny}{\zeta_{h_e}\bar{\gamma}_e}}\rb)}.
\end{align}

The normalized capacity per Hertz of bandwidth of the user channel and the eavesdropper channel are given by $C_\textrm{B}=\log_{2}(1+\gamma_d)$ and $C_\textrm{E}=\log_{2}(1+\gamma_e)$, respectively. The achievable secrecy rate ($C_\textrm{S}$) is then defined as \cite{Barros_2006_Secrecy-Capacity}
\begin{align}
\label{eq_sec_rate}
C_\textrm{S}&= [C_\textrm{B}-C_\textrm{E}]^+ =\max\lb[\log_2\lb(\frac{1+\gamma_d}{1+\gamma_e}\rb),0 \rb],
\end{align}
where $[x]^+=\max\{x,0\}$. 

\section{Secrecy Outage Probability (SOP) }

 \label{sec_SOP}
A secrecy outage happens when the instantaneous secrecy rate is less than the desired threshold secrecy rate $R_{\textrm{th}}$ for the system. The SOP can then be expressed  mathematically as \cite{Moualeu_2021_Physical-Layer-Security}
\begin{align}
&\mathcal{P}_{\textrm{out}}=\mathbb{P}\lb[\log_2\lb(\frac{1+\gamma_d}{1+\gamma_e}\rb)<R_\textrm{th} \rb],\nn\\
&=\mathbb{P}\left[\max_{\forall m}\{\lvert h_d^{(m)}\rvert^2\}\bar{\gamma}_d-\rho \max_{\forall n}\{\lvert h_e^{(n)}\rvert^2\} \bar{\gamma}_e\right.\left.<\frac{\lb(\rho-1\rb)}{\delta^2 \lvert g\rvert^2 }\right], \nn\\
\label{eq_sop4}
&=\mathbb{P}\left[\max_{\forall m}\{\lvert h_d^{(m)}\rvert^2\}\bar{\gamma}_d<\frac{\lb(\rho-1\rb)}{\delta^2\lvert g\rvert^2}\right. +\left. \rho \max_{\forall n}\{\lvert h_e^{(n)}\rvert^2\}\bar{\gamma}_e\right],
\end{align} 
where $\rho=2^{R_{\textrm{th}}}$.
To obtain the solution of the above expression, we first rewrite \eqref{eq_sop4} as 
\begin{align}
\label{eq_sop5}
\mathcal{P}_{\textrm{out}}&=\mathbb{P}\lb[X<\frac{(\rho-1)}{\delta^2 Z}+\rho Y\rb],
\end{align}
where $Z$ represents the RV $|g|^2$, and RVs $X$ and $Y$ are already defined in (\ref{cdf_X}) and (\ref{cdf_Y}), respectively. 
The expression of \eqref{eq_sop5} in the integral form can be written as 
\begin{align}
\label{Eq_sop_N_1_double_int}
\mathcal{P}_{\textrm{out}}
&=\int_{z=0}^{\infty}\int_{y=0}^{\infty}\lb[F_X\lb(\frac{(\rho-1)}{\delta^2 z}+\rho y\rb)\rb]f_{Y}(y)f_{Z}(z)dydz. 
\end{align}
By substituting the analytical expression of $F_{X}(x)$, $f_{X}(x)$, and $f_{Y}(y)$ already derived in \eqref{cdf_X}, \eqref{pdf_X}, and \eqref{pdf_Y}, respectively, the SOP is written as 
\begin{align}
\label{SOP-2}
&\mathcal{P}_{\textrm{out}}=\int_{z=0}^{\infty}\left[1+\sum^M_{m=1}\sum^N_{n=1}\binom{M}{m}\binom{N}{n}\frac{n\lb(-1\rb)^{m+n+1}}{\zeta_{h_e} \bar{\gamma}_e}\right.\nn \\
&\times \left.\int_{y=0}^{\infty}\exp{\lb({-\frac{m\lb(\frac{(\rho-1)}{\delta^2 Z }+\rho Y\rb)}{\zeta_{h_d}\bar{\gamma}_d}{-\frac{ny}{\zeta_{h_e}\bar{\gamma}_e}}}\rb)}dy\right]f_{Z}(z)dz.
\end{align}
After evaluating the inner integral and substituting the PDF $f_{Z}(z)$ from \eqref{pdf_Z} in (\ref{SOP-2}), we get
\begin{align}
\label{SOP-3}
\mathcal{P}_{\textrm{out}}&= 1-\sum^M_{m=1}\sum^N_{n=1}\binom{M}{m}\binom{N}{n}\frac{{(-1)^{m+n}}n\zeta_{h_d}\bar{\gamma}_d}{m\rho\zeta_{h_e}\bar{\gamma}_e+n\zeta_{h_d}\bar{\gamma}_d} \nn \\
& \times \underbrace{\int_{z=0}^{\infty}\frac{1}{\zeta_g}\exp{\lb(-\frac{m(\rho-1)}{\delta^2 z \zeta_{h_d}\bar{\gamma}_d}\rb)}\exp{\lb(-\frac{z}{\zeta_g}\rb)}dz}_{I}.
\end{align}
After some mathematical manipulations and with the aid of \cite[eq. (3.324.1)]{Gradshteyn_Ryzhik_2014_table}, 
which enables us to use the modified Bessel function for the solution of $I$, we get the closed-form solution for the SOP in (\ref{SOP-3}) as 
\begin{align}
\label{eq_sop_N_2_eval_final}
\mathcal{P}_{\textrm{out}}&=1-\sum^M_{m=1}\sum^N_{n=1}\binom{M}{m}\binom{N}{n}\frac{(-1)^{m+n}n\zeta_{h_d}\bar{\gamma}_d}{m\rho\zeta_{h_e}\bar{\gamma}_e+n\zeta_{h_d}\bar{\gamma}_d}\nn\\
&\times \sqrt{\frac{4m\lb(\rho-1\rb)}{\delta^2\zeta_{g}\zeta_{h_d}\bar{\gamma}_d}}{K}_1\lb(\sqrt{\frac{4m\lb(\rho-1\rb)}{\delta^2\zeta_{g}\zeta_{h_d}\bar{\gamma}_d}}\rb).
\end{align}

\section{Asymptotic Analysis}
 \label{sec_ASYM_SOP}
 The closed-form expression of the SOP derived in (\ref{eq_sop_N_2_eval_final}) does not provide useful insights into the system design. To address this issue, we derive a simple and explicit expression that approximates the SOP in the high-SNR regime. To this end, we assume that 
the transmit power $P$ increases asymptotically, i.e.  $P\rightarrow\infty$, leading to $\bar{\gamma}_d = \frac{P}{\sigma_d^2}\rightarrow\infty$.
Herewith we can use the following approximation in (\ref{eq_sop_N_2_eval_final}) \cite[eq. (9.6.9)]{abramowitz1972handbook}
\begin{equation} 
\label{eq_approxBessel}
z\lim_{z\rightarrow 0} K_{1}(z)=1,
\end{equation} 
which yields
\begin{align}
\label{eq_asymptotic}
    \lim_{\bar{\gamma}_d \to \ \infty }\sqrt{\frac{4m\lb(\rho-1\rb)}{\delta^2\zeta_{g}\zeta_{h_d}\bar{\gamma}_d}}{K}_1\lb(\sqrt{\frac{4m\lb(\rho-1\rb)}{\delta^2\zeta_{g}\zeta_{h_d}\bar{\gamma}_d}}\rb) = 1.
\end{align}
After substituting (\ref{eq_asymptotic}) in (\ref{eq_sop_N_2_eval_final}), the asymptotic SOP can be expressed as
\begin{align}
\label{eq_sop_final}
\mathcal{P}_{\textrm{out}}^{\infty}
&= 1-\sum^M_{m=1}\sum^N_{n=1}\binom{M}{m}\binom{N}{m} \frac{(-1)^{m+n}n\zeta_{h_d}\sigma_e^2}{m\rho\zeta_{h_e}\sigma_d^2+n\zeta_{h_d}\sigma_e^2} .
\end{align}

From (\ref{eq_sop_final}), it can be inferred that the asymptotic SOP saturates in the high-SNR depending on the number of users, the number of eavesdroppers, the channel parameters of the keyhole-to-users and the keyhole-to-eavesdroppers, the noise power at the users and the eavesdroppers, and the desired threshold secrecy rate. Furthermore, the asymptotic SOP does not depend on the scattering cross-section and the channel parameter of the source-to-keyhole channel.

\section{Numerical Results}
In this section, we evaluate the SOP and asymptotic SOP performance of the keyhole-aided multi-user communication system in the presence of multiple eavesdroppers. 
Unless otherwise specified, the system parameters are taken as ${\zeta_g} = 3$ dB, ${\zeta_{h_d}} = 6$ dB, ${\zeta_{h_e}} = -3$ dB, $R_{\textrm{th}} = 1$ bits per channel use (bpcu), and $\delta = 0.5$.

\label{sec_results}

\begin{figure}
 \centering
\includegraphics[width=2.6 in]{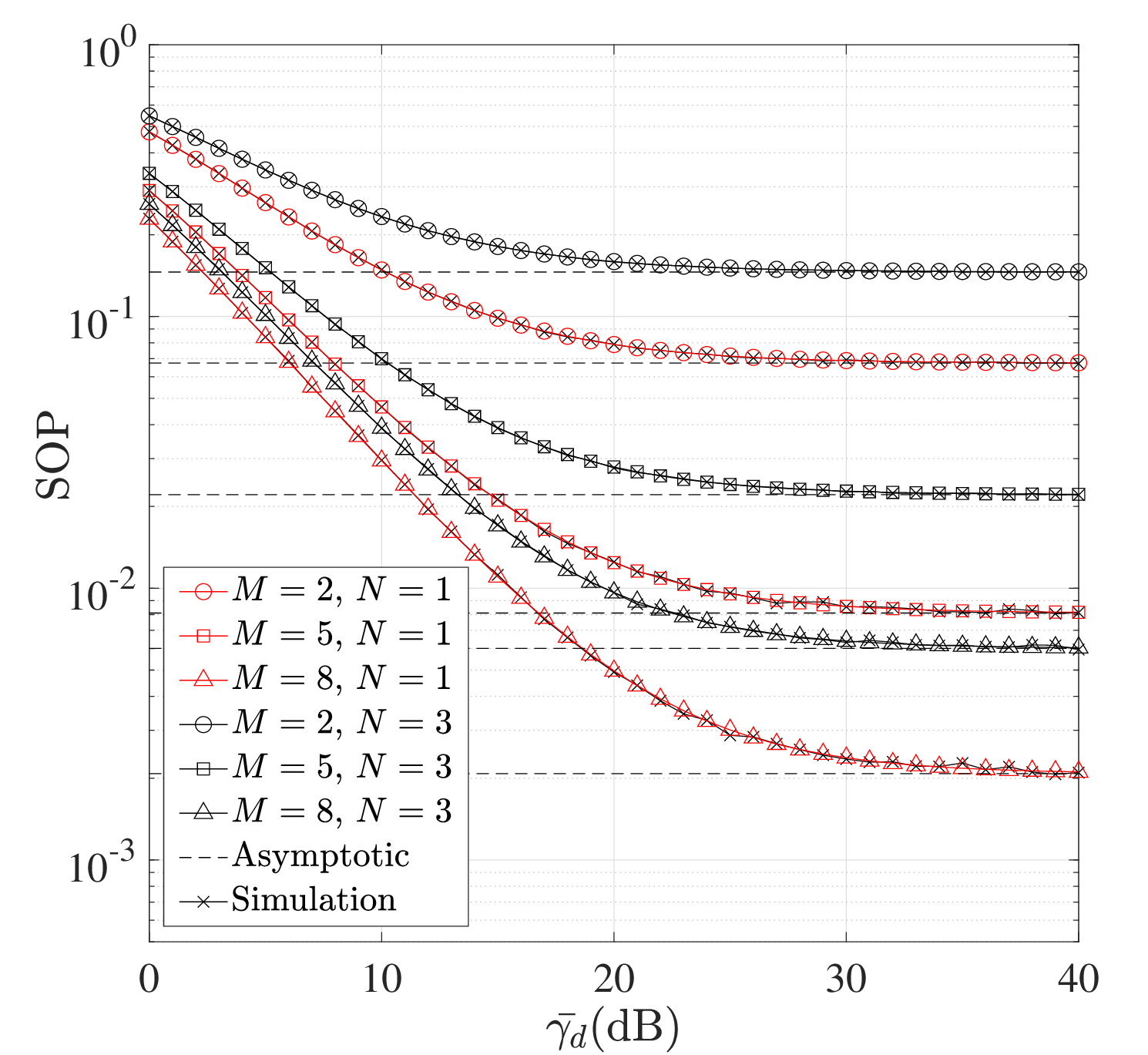} 
 \caption{The SOP versus $\bar{\gamma}_d$ for different $M$ and $N$.}
  \vspace{-.5cm}
 \label{fig_sop_vs_gamma_D_M_N_varying}
 \end{figure} 

 \begin{figure}
 \centering
\includegraphics[width=2.6 in]{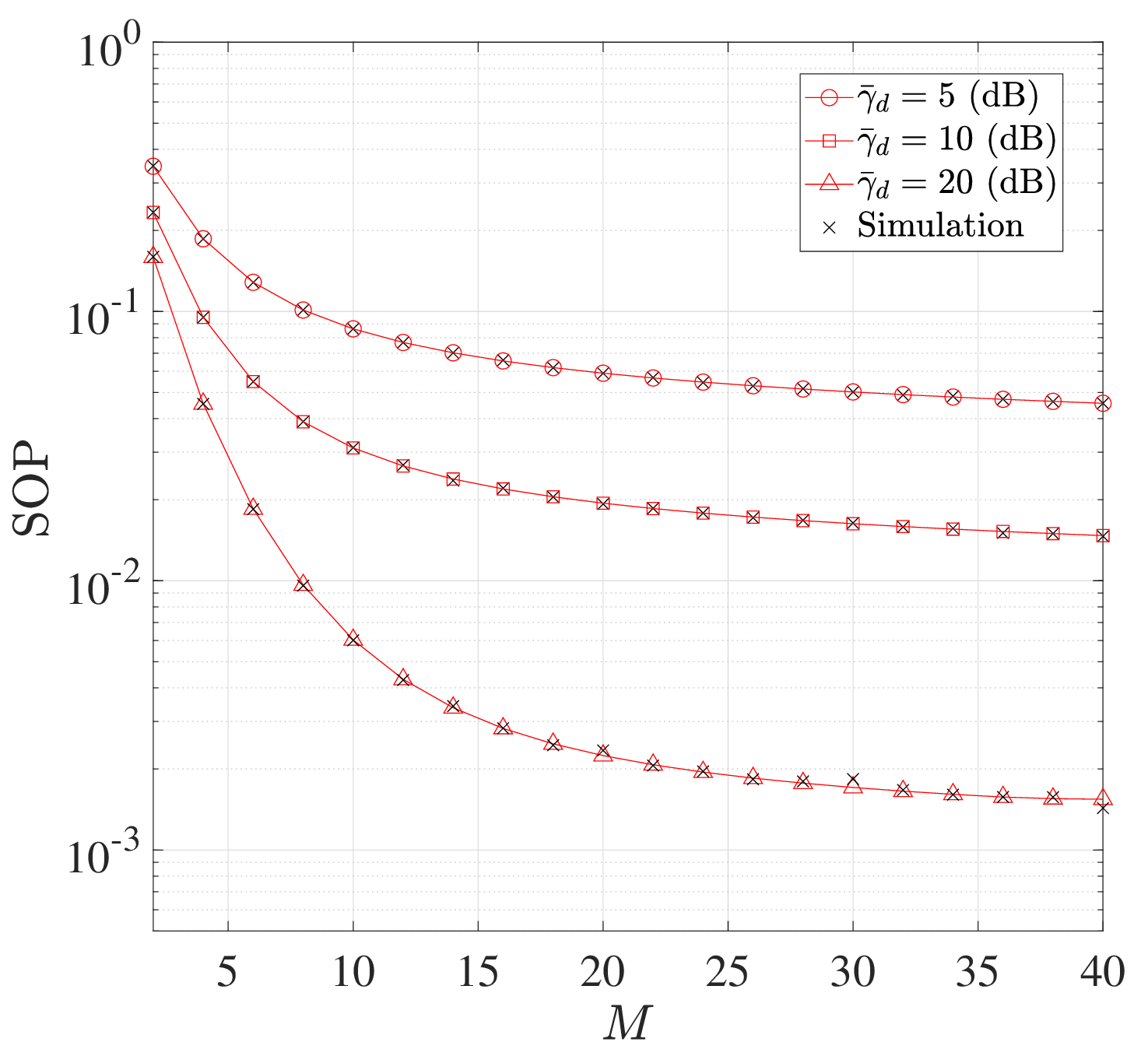} 
\caption{The SOP versus $M$ for different values of $\bar{\gamma}_d \in \{5,10,20\}$.}
 \label{fig_sop_vs_M_gamma_D_varying}
  \vspace{-.5cm}
 \end{figure}

\begin{figure}
 \centering
\includegraphics[width=2.6 in]{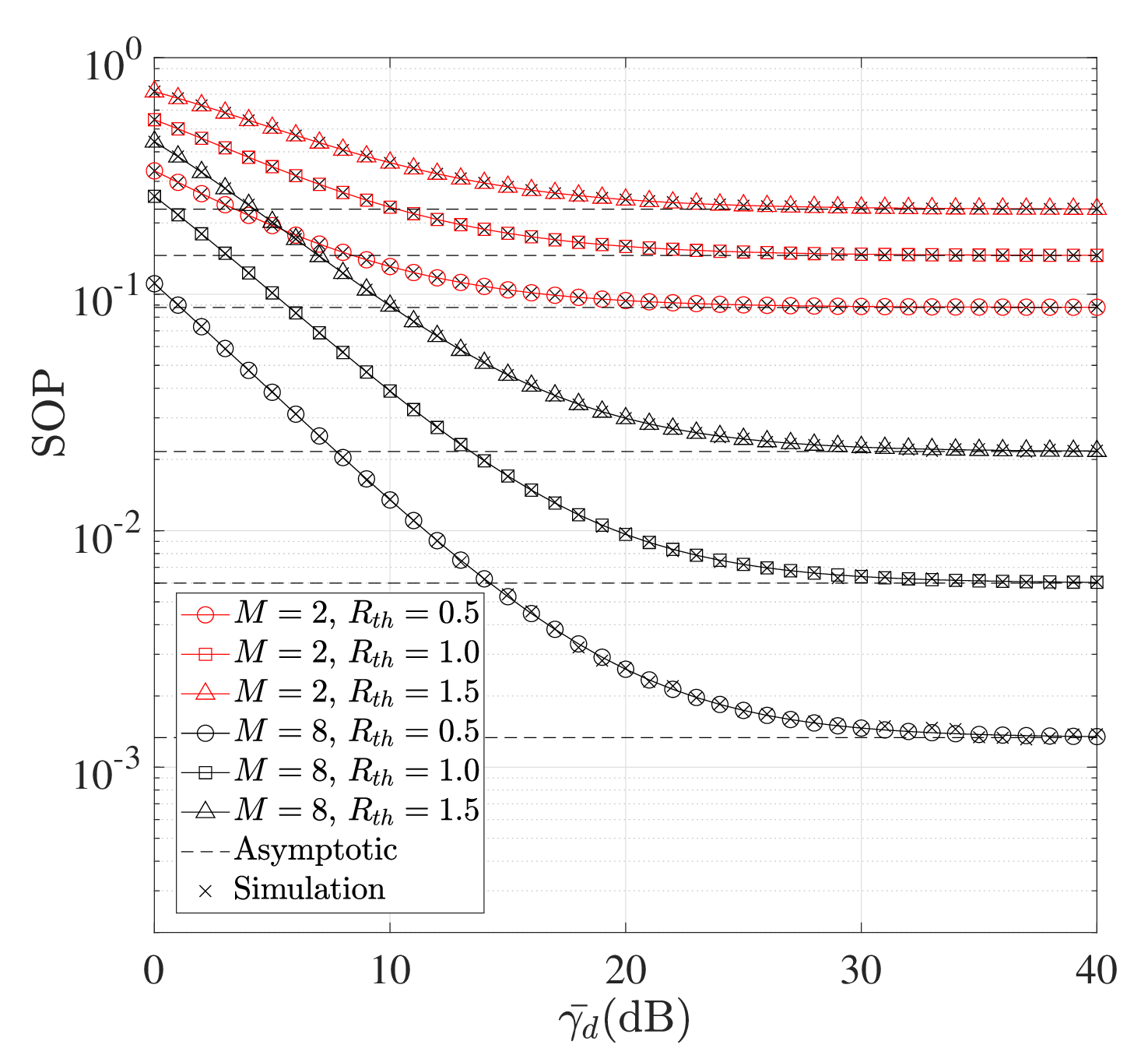} 
\caption{The SOP versus $\bar{\gamma}_d$ for $M \in \{2, 8 \}$ and $R_{\textrm{th}} \in \{0.5, 1, 1.5\}$.}
 \label{fig_sop_vs_MRvary}
  \vspace{-.5cm}
 \end{figure}

  In Fig. \ref{fig_sop_vs_gamma_D_M_N_varying}, the SOP is plotted against $\bar{\gamma}_d$ for the different combinations of $M \in \{2, 5, 8 \}$ and $N \in \{1, 3\}$. From the plots, it is observed that the theoretical SOP obtained in (\ref{eq_sop_N_2_eval_final}) perfectly matches with the corresponding Monte Carlo simulations, which validates the correctness of our analysis. In general, we observe that the SOP improves with the increase in $\bar{\gamma}_d$ and $M$ and with the decrease in $N$. For example, among the considered parameter combinations, the best performance is observed when the highest value of $M$ and the lowest value of $N$ are considered, i.e., when $M = 8$ and $N = 1$. This is because the diversity in the keyhole-to-user channel improves as $M$ increases, leading to improved user channel SNR. On the contrary, diversity reduces in the keyhole-to-eavesdropping channel as $N$ decreases. We also observe that the SOP improves with $\bar{\gamma}_d$ in the low-SNR regime; however, it saturates in the high-SNR regime. The saturation value exactly matches the asymptotic SOP derived in (\ref{eq_asymptotic}) in the high-SNR regime.

In Fig. \ref{fig_sop_vs_M_gamma_D_varying}, the SOP is plotted against $M$ to see the effect of $M$ on the SOP for the different values of $\bar{\gamma}_d \in \{5, 10, 20 \}$ dB when  $N = 3$. The SOP performance improves sharply with the increase in $M$ initially, however, a diminishing return can be seen as $M$ increases. As $M$ increases, the rate of performance improvement gradually decreases and at higher values of $M$, the performance gradually saturates depending on other system parameters.

 Fig. \ref{fig_sop_vs_MRvary} shows the impact of the threshold secrecy rate $R_\textrm{th}$ on the SOP for the different combinations of $M \in \{2, 8 \}$ and $N = 3$. 
 As the required threshold $R_{\textrm{th}}$ increases, the SOP performance decreases for a given $M$, which is intuitive. At a higher value of $M$,  changes in $R_{\textrm{th}}$ impact the SOP performance more as the performance gaps corresponding to different $R_{\textrm{th}}$ are more when $M = 8$ than when $M = 2$.
\begin{figure}
 \centering
\includegraphics[width=2.6 in]{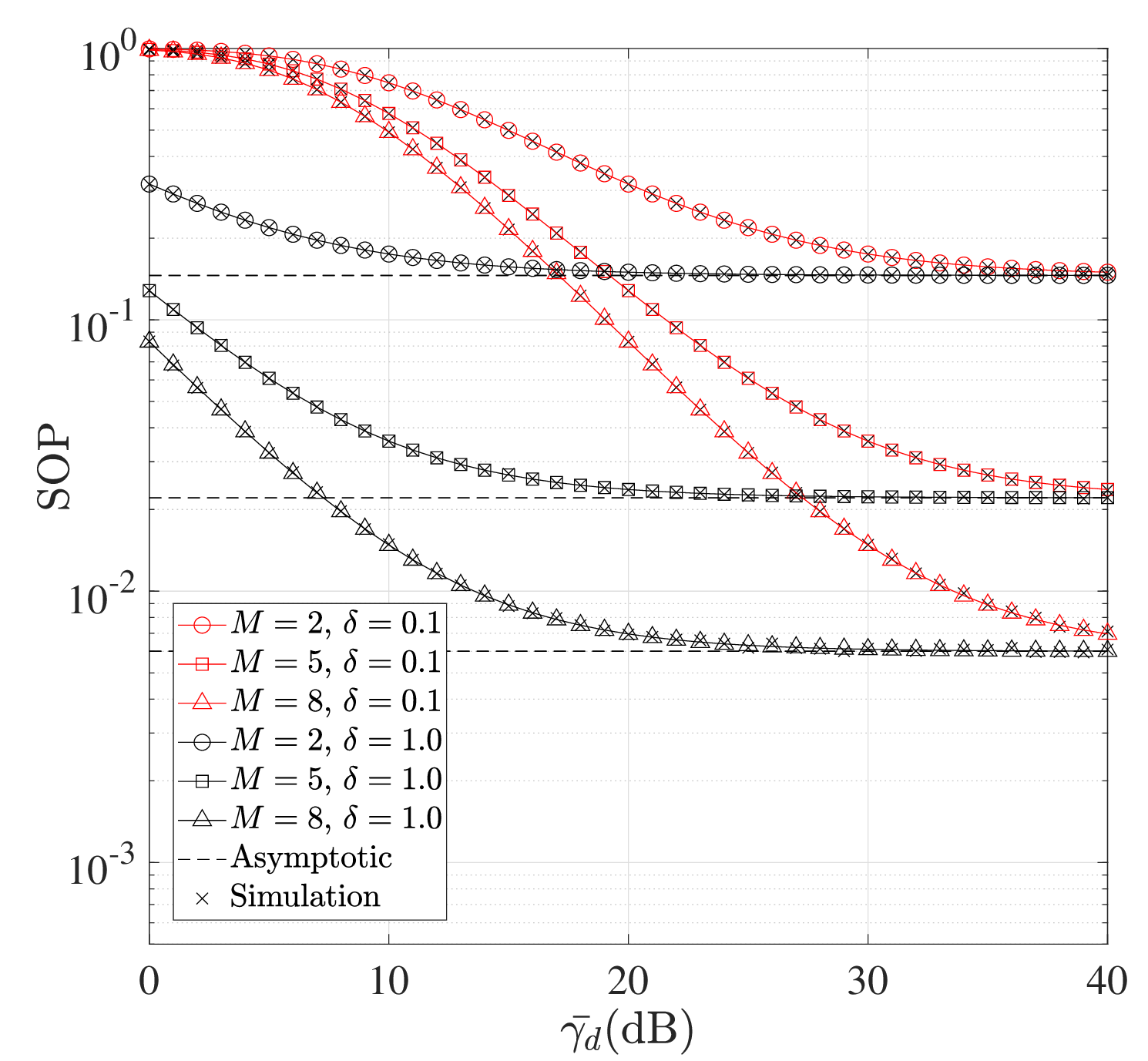}
\caption{The SOP versus $\bar{\gamma}_d$ for $M \in \{2, 5, 8 \}$ and $\delta\in \{0.1, 1.0 \}$.}
 \label{fig_sop_vs_SNR_delta_vary}
 \vspace{-.5cm}
 \end{figure}

Fig. \ref{fig_sop_vs_SNR_delta_vary} shows the impact of the keyhole parameter $\delta$ on the SOP for the different combinations of $M \in \{2, 5, 8 \}$ and $N=3$. We observe that the SOP improves for the higher values of $\delta$ for any $M$. The higher value of $\delta$ represents a large keyhole size, which leads to better transmission with stronger multi-path propagation. For example, we see that the SOPs when $\delta = 1$ for any $M$ is far better than that of $\delta = 0.1$ because the small $\delta$ results in severe signal attenuation and deep fades. In addition, we observe no influence of $\delta$ on the SOP performance in the high-SNR regime, i.e., the SOPs for $\delta = 0.1$ and $\delta = 1$ provide the same asymptotic results for a particular $M$ in the high-SNR regime. This corroborates the remark made after obtaining the asymptotic SOP in (\ref{eq_sop_final}), the asymptotic expression is independent of $\delta$ in the high-SNR regime.
\vspace{-.2cm}
\section{Conclusions}
 \label{sec_conclusion}
In this paper, we have presented the secrecy performance analysis
of a single keyhole-aided multi-user communication network in the presence of multiple eavesdroppers. 
Wherein the wireless link experiences an independent Rayleigh fading.
The keyhole supports the transmission of both legitimate and unauthorized links. We have derived the exact closed-form expression of the SOP and asymptotic SOP of a user scheduling technique for the keyhole-aided network. Numerical results have shown that the asymptotic SOP in the high-SNR regime saturates to a constant value depending on the system parameters. As such, it can be inferred that there is no influence of the scattering cross-section parameter of the keyhole and the channel parameter of the source-to-keyhole channel on the SOP performance.
\section*{Acknowledgment}
This work was supported in part by the Ministry of Electronics and Information Technology (MeitY), Govt. of India through its project SwaYaan-Capacity Building in Drone/UAS with Ref. No. L14011/29/2021-HRD and in part by
Taighde Éireann – Research Ireland under Grant number 22/PATH-S/10788.



\bibliographystyle{IEEEtran}
\bibliography{IEEEabrv,references} 

\end{document}